# Impeding effect of cerium on the growth of helium bubble in iron

W. Hao and W. T. Geng[a]

*School of Materials Science and Engineering, University of Science and Technology Beijing, Beijing 100083, China*

Our first-principles density functional theory calculations suggest that the rare earth element, Ce, has a stronger attraction (-1.31eV) to He than He-He (-1.18eV) in bcc Fe. Consequently, the mobile He atoms could be pinned to Ce, and hence a reduced merging of He clusters. Moreover, we find that the segregated Ce layer at the He bubble surface presents an energy barrier of 0.33 eV to the upcoming He atom and thus slows down the bubble growth.

---

[a] To whom correspondence should be addressed. E-mail: geng@ustb.edu.cn

He bubbles formed during irradiation in nuclear reactor materials have been presenting a big challenge for material researchers since the application of nuclear energy [1]. As candidate structure materials in fusion reactor, the properties of low-activation martensitic steels (bcc Fe based alloy) under have been intensively studied both experimentally and theoretically [2, 3]. Problems induced by He bubble, embrittlement for instance, are fatal to the material's mechanical properties, particularly at high temperatures. Fu *et al.*'s calculations have showed that the diffusion barrier for interstitial helium in Fe is only 0.06 eV [4]. Such a low barrier nearly allows He atom to go anywhere they prefer. What's more, due to their strong self-trapping [5] and strong binding to vacancies [6] and grain boundaries [7], the accumulation of He atoms in Fe is practically unpreventable. To fight against with He embrittlement, we need to hinder both the growth and movement (or merge) of He bubbles. In our previous work, we predicted that gold could slow down He bubble growth in bcc Fe due to the stronger bonding between Au-Au than Fe-Fe and Au-Fe at the bubble surface [8]. Obviously, the shortcoming of using Au to prevent He bubble growth is the high cost. But the idea of caging He bubbles by alloy element to prevent bubble growth is valuable and instructive for designing fusion reactor materials. According to this idea, we searched another candidate element Ce to slow down He bubble growth. Cerium is the most abundant of the rare earth elements, making up about 0.0046% of the Earth's crust by weight. Because of the high affinity of cerium to sulfur and oxygen, it is used in various iron alloys to help reduce sulfides and oxides, and it is a precipitation hardening agent in stainless steel [9]. In this paper, we have investigated the interactions of Ce with He atoms and the retarding effect of Cr on He bubble growth by means of the first-principles calculations.

Our calculations were performed within the density functional theory (DFT) [10] as implemented within the Vienna *ab initio* simulation package (VASP) [11]. The electron-ion interaction and the exchange correlation between electrons were respectively described by the projector augmented wave (PAW) method [12] and the generalized gradient approximation (GGA) in the Perdew-Burke-Ernzerhof (PBE) form [13]. The cutoff energy for the plane-wave basis was set 480 eV, and a ($2\times2\times2$) *k*-mesh within Monkhorst-Pack scheme [14] was used. The volume of the supercell was fixed but all the internal freedoms were fully relaxed and the geometries were optimized until the forces on all the atoms were converged to less than $10^{-3}$ eV Å$^{-1}$. The calculated lattice constant for bcc Fe was 2.83 Å, which was yielded by our GGA-PBE computation.

The interaction between Ce and He in bcc Fe was evaluated within a (4×4×4) bcc Fe supercell (Figure 1). We find that He and Ce attract each other strongly and the binding energy, defined as the energy change of the supercell when the two move to neighboring sites from far apart, is as large as -1.31 eV. Note that, in bcc Fe the calculated binding energy of two He atoms is -1.18 eV [15]. That is to say, a He atom is more willing to cluster with Ce than He itself. Based on this calculation result, we believe that He would firstly pin to Ce atoms other than forming He cluster if Ce is alloyed into iron. Thus the He atoms can be discretely pinned to Ce instead of forming He clusters, which are the predecessor of bubble. On the other hand, according to the fact that Ce is attracted to an individual He atom we argue that Ce will also segregate to the He bubble surface. Differently from Au, Ce atoms can't form a cubic cage for He due to its bigger atom radius. However, a tetrahedron cage of Ce for He is energitically permissible. Following the numerical order marked in Fig 1, we replaced the Fe atoms around the substitutional He atom using Ce atom one by one. The calculated trapping energies are listed in Table I. From Table I, we can clearly see that all the four Ce atoms have a strong binding energy to He. So we have theoretically confirmed that a segregated Ce wall will be formed outside the He bubble.

**Fig. 1 (Color Online) A tetrahedron Ce cage encapsulating a He atom in bcc Fe. The numbers stand for the adding order of Ce atoms.**

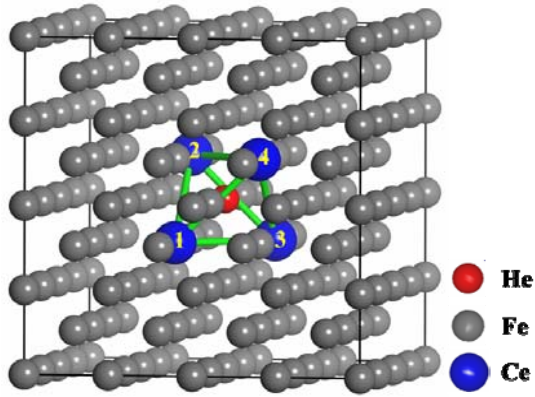

**Table I. The calculated energy change when one Ce atom is attracted from a perfect bcc Fe bulk environment to the vicinity of a substitutional He. The trapping process is in a sequential manner.**

| $n$ | 1 | 2 | 3 | 4 |
|---|---|---|---|---|
| $\Delta E(n)$ /eV | -1.31 | -0.88 | -0.65 | -0.55 |

**Fig. 2 (Color online) Energy barrier for the upcoming interstitial He in bcc Fe to enter the Ce cage (see Figure 1).**

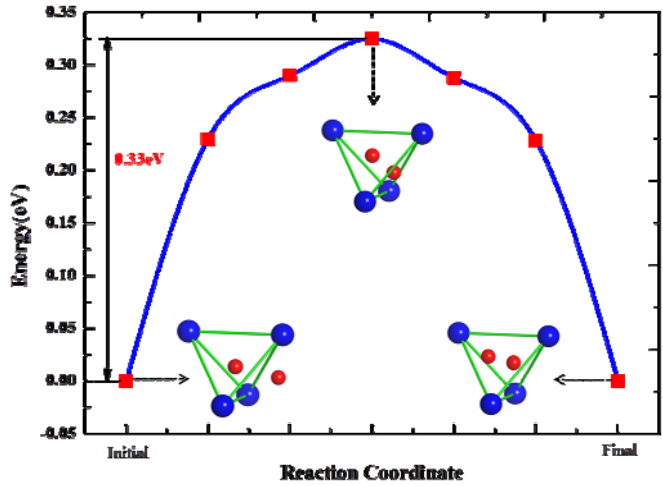

Then, the diffusion barrier for a He atom in Fe to go through the Ce wall and join the encapsulated He was calculated using the climbing nudged elastic band (CNEB) method [16]. The result is displayed in Figure 2. The upcoming He atom must overcome an energy barrier of 0.33 eV to join the caged He. While for the case of Fe cage, the barrier is essentially zero, agreeable with the fact that the diffusion barrier for interstitial He in bcc Fe is as small as 0.06 eV. Although the energy barrier of 0.33 eV is just about one half of the Au case, we can still expect that the growth of He bubble will be slowed down.

In order to understand the effect of segregated Ce on the growth of He bubble, we analysed the distrubition of valence electrons in (111) plane containing Ce1-Ce3 of Fig. 1. As plotted in Fig. 3, we find that in this plane a higher charge density exists between Ce-Ce atoms, which indicate that the Ce-Ce bonding is stronger than Ce-Fe and Fe-Fe bonding. Since the 1s orbital of He does not hybridize with orbitals of any other atoms, it prefers in general to stay in a low-electron density position. As a result, it will be more difficult for an upcoming He to pass through the Au cage than an Fe cage.

**Fig. 3 Distribution of valence electrons in the (111) plane containing Ce1-Ce3 of Fig. 1. Lines start from 0.025 $e$/a.u.$^3$ and increase successively by a factor of $10^{1/5}$.**

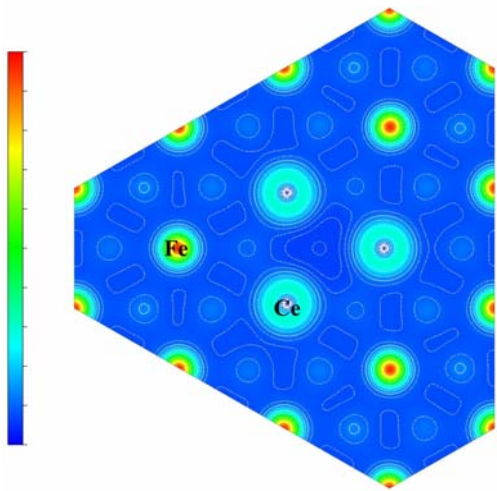

In summary, we predict by first-principles calculations that Ce has a stronger attraction to He than He atom itself in bcc Fe. Thus instead of forming He cluster, He atoms will be firstly pinned to Ce atom, making the distribution of He atoms dispersed. On the other hand, the segregated Ce on helium bubble performs a retarding effect on bubble growth. Although the effect of Ce is not as significant as that of Au, the cost of materials is much acceptable. Meanwhile, experimental scrutiny of our first-principles prediction is called for.

**Acknowledgments**

We are grateful to the support of the NSFC (Grant No. 50971029) and MOST (Grant No. 2009GB109004) of China, and NSFC-ANR (Grant No. 51061130558).